\documentclass[twocolumn,showpacs,preprintnumbers,amsmath,amssymb]{revtex4}

\usepackage{graphicx}
\usepackage{dcolumn}
\usepackage{bm}

\begin{document}


\title {The equations of motion
for consistency of a post-Newtonian Lagrangian formulation
} 
\author{Dan Li}
\author{Yu Wang} 
\author{Chen Deng} 
\author{Xin Wu}
\email{xinwu@gxu.edu.cn} \affiliation{School of Physical Science
and Technology, Guangxi University, Nanning 530004, China}


\begin{abstract}

Equations of motion for a general relativistic post-Newtonian
Lagrangian approach mainly refer to acceleration equations, i.e.
differential  equations of velocities. They are directly from  the
Euler-Lagrangian equations, and usually have higher-order terms
truncated when they remain at the same post-Newtonian order of the
Lagrangian. In this sense, they are incoherent equations of the
Lagrangian and approximately conserve constants of motion in this
system. In this paper, we show that the Euler-Lagrangian equations
can also yield the equations of motion for consistency of the
Lagrangian in the general case. The coherent equations are the
\emph{differential} equations of generalized momenta rather than
those of the velocities, and have no terms truncated. The
velocities are not integration variables, but they can be solved
from the \emph{algebraic} equations of the generalized momenta
with an iterative method. Taking weak relativistic fields in the
Solar System and strong relativistic fields of compact objects as
examples, we numerically evaluate the accuracies of the constants
of motion in the two sets of equations of motion. It is confirmed
that these accuracies well satisfy the theoretical need if the
chosen integrator  can provide a high enough precision. The
differences in the dynamical behavior of order and chaos  between
the two sets of equations are also compared. Unlike the incoherent
post-Newtonian Lagrangian equations of motion, the coherent ones
can theoretically, strictly conserve all integrals in some
post-Newtonian Lagrangian problems, and therefore are worth
recommending.

\end{abstract}


\maketitle

\section{Introduction}

The post-Newtonian (PN) Lagrangian approximations and the PN
Hamiltonian approximations are often visible in relativistic
astrophysics. They are widely applied to the description of the
equations of motion of generic black hole binaries at a certain
high PN order [1-3], including spin corrections. In this way,
high-precision theoretical templates of gravitational waveforms
can be provided. The PN approximations are also used to treat the
equations of motion for the relativistic restricted three-body
problem [4] and those for the relativistic $N$-body gravitational
problem [5]. When one of the two PN formalisms is converted to
another at the same order, their physical equivalence was shown in
Refs. [6-8]. Due to higher-order terms truncated, the two
formulations have somewhat differences  and  are not exactly equal
[9-11]. For the Solar System as a weak gravitational field, the
differences are too small to affect their equivalence, that is,
the solutions of the two formulations should have no typical
differences for the regular case. Some famous PN effects like
perihelion or periastron advances for spinless binaries or the
geodetic, the Lense-Thirring and the Schiff precessions for
spinning bodies are the same in the two formulations. However, the
differences lead to the two formulations having different
solutions during a long integration time for the ordered case in a
strong gravitational field of compact objects. Sometimes the
differences even would make the two formulations have different
dynamical properties of integrability and nonintegrability, or
order and chaos.

It is well known that the paths for deriving the PN Hamiltonian
equations of motion and the PN Lagrangian equations of motion are
different. The Hamiltonian equations of motion, i.e. the canonical
equations of a Hamiltonian formulation, have no terms truncated.
Therefore, the equations of motion are consistent with this PN
Hamiltonian. In this case, the constants of motion such as the
energy integral are strictly conserved by the PN Hamiltonian
equations of motion. The PN Lagrangian equations of motion are
associated to the acceleration equations that result from the
Euler-Lagrangian equations of a PN Lagrangian formulation. If the
generalized momenta given by the PN Lagrangian system are
nonlinear functions of the velocities, then the accelerations
appear as the PN terms of the Euler-Lagrangian equations. When the
total accelerations are required to remain at the same PN order of
the Lagrangian, the accelerations in the PN terms of the
Euler-Lagrangian equations must be replaced with the lower-order
equations of motion. The higher-order PN terms truncated in the
Euler-Lagrangian equations make the acceleration equations
incoherent. That is, the acceleration equations are inconsistent
with this PN Lagrangian formulation. Thus, the constants of motion
are not strictly conserved by the incoherent acceleration
equations. If the Lagrangian equations of motion can be given
coherently, then they can naturally, strictly conserve the
constants of motion in the PN Lagrangian formulation. Without
doubt, both the incoherent Lagrangian equations of motion and the
coherent ones have somewhat differences. When these differences
are properly large, the two sets of equations of motion may
exhibit distinct dynamical behaviors.

There is a question of whether the coherent Lagrangian equations
of motion can be written. Recently, the authors of [12] suggested
that coherent implicit  acceleration equations, derived from the
Lagrangian of a PN circular restricted three-body problem [4, 13],
should be integrated by using implicit numerical integrators.
Unlike them, we provide a simple method to construct the coherent
Lagrangian equations of motion that strictly conserve the
constants of motion. In our method, the differential equations of
the generalized momenta rather than those of the velocities (i.e.
the acceleration equations) directly result from the
Euler-Lagrangian equations. They are still solved with an explicit
numerical integrator. However, the velocities as not integration
variables must be solved from the algebraic equations of the
generalized momenta by means of an iterative method.

The paper is organized as follows. In Sect. 2, we provide two sets
of PN equations of motion for a given PN Lagrangian. To clearly
show the difference between them, we list two examples in detail.
The PN circular restricted three-body problems are considered in
Sect. 3. Then, PN Lagrangian systems of compact binaries are
tested in Sect. 4. Finally, our main results are concluded in
Sect. 5.

\section{PN Lagrangian equations of motion}

Let us consider a PN Lagrangian
$\mathcal{L}(\mathbf{r},\mathbf{v})$ at $m$th order, where
$\mathbf{r}$ is a position vector and $\mathbf{v}$ is a velocity
vector. We have the generalized momentum
\begin{equation}
\mathbf{p}=\frac{\partial \mathcal{L}}{\partial \mathbf{v}}.
\end{equation}
In general, $\mathbf{p}$ is a nonlinear \emph{algebraic} equation
of  $\mathbf{v}$. The Euler-Lagrangian equation is
\begin{equation}
\frac{d\mathbf{p}}{dt}=\frac{\partial \mathcal{L}}{\partial
\mathbf{r}}.
\end{equation}
Obviously, this equation is a  \emph{differential} equation of the
momentum $\mathbf{p}$ to time $t$ and remains at the same PN order
of the PN Lagrangian $\mathcal{L}$. In the following discussions,
we use two methods to derive the equations of motion from the PN
Lagrangian.

\subsection{Two sets of PN equations of motion}

Besides Eq. (2), another equation is added by
\begin{equation}
\frac{d\mathbf{r}}{dt}=\mathbf{v}.
\end{equation}
When we solve  the differential equations (2) and (3), the pair
$(\mathbf{r},\mathbf{p})$ is viewed as a set of integration
variables, but the velocity $\mathbf{v}$ is not. In spite of this,
the velocity can be solved from the generalized momentum algebraic
equation (1) using  an iterative method like the Newtonian
iterative method. It is clear that no truncations occur when Eqs.
(2) and (3) are obtained from the PN Lagrangian. Therefore, Eqs.
(2) and (3) with Eq. (1) are called as coherent equations of
motion for the PN Lagrangian $\mathcal{L}$. Without doubt, they
strictly conserve constants of motion, such as the energy integral
\begin{equation}
E=\mathbf{v}\cdot\mathbf{p}-\mathcal{L}.
\end{equation}

On the other hand, one usually substitutes Eq. (1) into Eq. (2)
and obtains the  acceleration
\begin{equation}
\frac{d\mathbf{v}}{dt}=\mathbf{a}_{N}+\mathbf{a}_{1PN}+\cdots+\mathbf{a}_{mPN},
\end{equation}
where $\mathbf{a}_{N}$, $\mathbf{a}_{1PN}$, $\cdots$,
$\mathbf{a}_{mPN}$ represent the  accelerations from the Newtonian
term, first post-Newtonian order contribution, $\cdots$, $m$th
post-Newtonian order contribution. Since the total acceleration
$d\mathbf{v}/dt$ remains at $m$th PN order, all terms higher than
the order $m$ must be truncated. This shows that Eq. (5) is
inconsistent with the PN Lagrangian $\mathcal{L}$. That is to say,
Eqs. (3) and (5) use $(\mathbf{r},\mathbf{v})$ as a set of
integration variables and are incoherent PN equations of motion
for the Lagrangian $\mathcal{L}$. Naturally, they approximately
conserve the energy integral (4).

It is worth noting that the iterative accuracy of the velocity
$\mathbf{v}$ in Eq. (1) reaches an order of $10^{-15}$, which
almost approaches to the double precision of the machine, an order
of $10^{-16}$. This accuracy is much higher than the accuracy of
the PN terms truncated in strong gravitational problems. This is
why Eqs. (1)-(3) are called as the coherent equations of motion,
and Eqs. (3) and (5), the incoherent equations of motion.
Additionally, the incoherent equations of motion and the coherent
ones are not completely the same but are only approximately
related two different dynamical systems.

\subsection{Comparison with PN Hamilton's equations}

An $m$th PN order Hamiltonian $\mathcal{H}(\mathbf{r},\mathbf{p})$
(nonequivalent to $\mathcal{L}$) has canonical  equations
\begin{eqnarray}
\frac{d\mathbf{r}}{dt} &=& \frac{\partial \mathcal{H}}{\partial \mathbf{p}}, \\
\frac{d\mathbf{p}}{dt} &=& -\frac{\partial \mathcal{H}}{\partial
\mathbf{r}}.
\end{eqnarray}
The canonical  equations are coherent equations of motion for this
PN Hamiltonian, and exactly conserve the Hamiltonian.

By comparing between the coherent PN Lagrangian equations (2) and
(3) and the coherent PN Hamilton's equations (6) and (7), one can
easily find that they are very similar. They are the differential
equations with respect to $\mathbf{r}$ and $\mathbf{p}$, and
remain the same PN order of $\mathcal{L}$ or $\mathcal{H}$.
However, there are some typical differences between them. That is,
the velocity must be calculated via  a certain iterative method in
the coherent PN Lagrangian equations, but does not need such a
calculation in the coherent PN Hamilton's equations. In addition
to this, $\mathbf{r}$ and $\mathbf{p}$ are not canonical in the
coherent PN Lagrangian equations, but are in the coherent PN
Hamilton's equations.

Perhaps someone may think that the  acceleration in the coherent
PN Lagrangian equations should be higher than the order $m$ when
Eq. (1) is substituted  into Eq. (2). It is true. In fact, the
acceleration, obtained from the coherent PN Hamilton's equations
by substituting $\mathbf{p}$ (given by Eq. (6)) into Eq. (7), is
also higher than  the order $m$.

It is clear that  the coherent PN Lagrangian equations of motion
and the incoherent ones exist some differences in higher-order
terms. Do they show different dynamical behaviors? To answer this
question, we compare them using numerical simulations of three
problems.

\section{PN circular restricted three-body problems}

In  a planar circular restricted three-body problem, two primary
bodies with masses $M_1$ and $M_2$ always have an invariant
separation $a$ because they are restricted to moving  in circular
orbits around their center of mass under their gravities. A third
body has such a small mass $m$ that  it does not exert any
influence on the circular motions of the binaries. Let $M=M_1+M_2$
be the total mass of the primary bodies, and take the
gravitational constant as a geometrized unit $G=1$. When  the
distances (e.g. $a$) and time $t$ are measured in terms of $M$,
the two primaries have dimensionless masses $\mu_1=M_1/M=1-\mu$
and $\mu_2=M_2/M=\mu$. If we further adopt the unit system of
[13], that is,  the distances and time  are respectively measured
in terms of $a$ and $1/\omega_0$ with the Newtonian angular speed
of the primaries $\omega_0=a^{-3/2}$, then the distance of the
binaries and the Newtonian angular speed of the circular orbits of
the binaries become one geometric unit. In this case, the two
central bodies are fixed at points $(-\mu,0)$ and $(1-\mu,0)$ in
the rotating frame. Under the gravities of the binaries, the
motion of the third body with coordinate $\textbf{r}=(x,y)$ and
velocity $\textbf{v}=(\dot{x},\dot{y})$ is described by the
following  dimensionless 1PN  Lagrangian formalism
\begin{equation}
\mathcal{L}=\mathcal{L}_0+\frac{1}{c^{2}}(\mathcal{L}_1+\mathcal{L}_2),
\end{equation}
where $c$ is the velocity of light. The first part is the
Newtonian circular restricted three-body problem
\begin{eqnarray}
\mathcal{L}_0 &=&
\frac{1}{2}(\dot{x}^{2}+\dot{y}^{2})+x\dot{y}-\dot{x}y
+\frac{1}{2}(x^{2}+y^{2}) \nonumber \\
&&  +\frac{1-\mu}{r_1}+\frac{\mu}{r_2}, \\
r_1 &=& \sqrt{(x+\mu)^{2}+y^{2}}, \nonumber \\
r_2 &=& \sqrt{(x-1+\mu)^{2}+y^{2}}. \nonumber
\end{eqnarray}
The second part is an indirect 1PN  relativistic effect to the
circular orbits of the two central bodies, which affects the
motion of the third body. It is expressed as
\begin{equation}
\mathcal{L}_1 = \omega_1(x\dot{y}-\dot{x}y +x^{2}+y^{2}),
\end{equation}
where $\omega_1$ is the PN angular velocity of  the primaries
\begin{equation}
\omega_1 = [(1-\mu)\mu-3]/(2a).
\end{equation}
The third part is a direct 1PN  relativistic effect to the third
body. For our purpose, we choose a part of the complete expression
on the relativistic effect [4, 13]:
\begin{eqnarray}
\mathcal{L}_2 &=&
\frac{1}{8a}[\dot{x}^{2}+\dot{y}^{2}+2(x\dot{y}-\dot{x}y)
+x^{2}+y^{2}]^{2} \nonumber \\
&&  +\frac{3}{2a}(\frac{1-\mu}{r_1}+\frac{\mu}{r_2})[\dot{x}^{2}+\dot{y}^{2}+2(x\dot{y}-\dot{x}y) \nonumber \\
&& +x^{2}+y^{2}+\mu(1-\mu)].
\end{eqnarray}

Noting Eq. (1), we have the generalized momenta
\begin{eqnarray}
p_x &=&
\dot{x}-y-\frac{\omega_1}{c^{2}}y+\frac{\dot{x}-y}{2ac^{2}}
[\dot{x}^{2}+\dot{y}^{2}  \nonumber \\
& &  +2(x\dot{y}-\dot{x}y) +x^{2}+y^{2}  \nonumber \\
& &  +6(\frac{1-\mu}{r_1}+\frac{\mu}{r_2})], \\
p_y &=&
\dot{y}+x+\frac{\omega_1}{c^{2}}x+\frac{\dot{y}+x}{2ac^{2}}
[\dot{x}^{2}+\dot{y}^{2}  \nonumber \\
& &  +2(x\dot{y}-\dot{x}y) +x^{2}+y^{2}  \nonumber \\
& &  +6(\frac{1-\mu}{r_1}+\frac{\mu}{r_2})].
\end{eqnarray}
Based on Eq. (2), the Euler-Lagrangian equations are
\begin{eqnarray}
\frac{dp_x}{dt} &=& \frac{\partial \mathcal{L}}{\partial x}=\mathcal{L}_{x}, \\
\frac{dp_y}{dt} &=& \frac{\partial \mathcal{L}}{\partial
y}=\mathcal{L}_{y}.
\end{eqnarray}
Eq. (3) corresponds to the following two equations
\begin{eqnarray}
\dot{x} &=& v_{x}, \\
\dot{y} &=& v_{y}.
\end{eqnarray}
Eqs. (15)-(18) are the above-mentioned coherent equations of
motion for the Lagrangian (8). In terms of Eqs. (13) and (14), we
use  an iterative method to get the velocities
\begin{eqnarray}
v_x &=& \{p_x+y[1+\frac{\omega_1}{c^{2}}+\frac{1}{2ac^{2}}
((x+v_y)^{2} \nonumber \\
& &  +(v_x-y)^{2}+6(\frac{1-\mu}{r_1}+\frac{\mu}{r_2}))]\}\nonumber \\
& &  /\{1+\frac{1}{2ac^{2}}
[(x+v_y)^{2}+(v_x-y)^{2} \nonumber \\
&& +6(\frac{1-\mu}{r_1}+\frac{\mu}{r_2})]\}, \\
v_y &=& \{p_y-x[1+\frac{\omega_1}{c^{2}}+\frac{1}{2ac^{2}}
((x+v_y)^{2} \nonumber \\
& &  +(v_x-y)^{2}+6(\frac{1-\mu}{r_1}+\frac{\mu}{r_2}))]\}\nonumber \\
& &  /\{1+\frac{1}{2ac^{2}}
[(x+v_y)^{2}+(v_x-y)^{2} \nonumber \\
&& +6(\frac{1-\mu}{r_1}+\frac{\mu}{r_2})]\}.
\end{eqnarray}
As a suitable choice of the initial values of $(v_x,v_y)$ in the
right sides of the above equations, $v_x=p_x$ and $v_y=p_y$ are
suggested. It is clear that the denominators in Eqs. (19) and (20)
are larger than 1. Therefore, the iterative solutions are
convergent.  Eqs. (15)-(18) with Eqs. (19) and (20) strictly
conserve the energy
\begin{eqnarray}
E &=& \dot{x}p_x+\dot{y}p_y-\mathcal{L} \nonumber \\
&=&-\frac{C_J}{2}-(1+\frac{\omega_1}{c^{2}})(x^{2}+y^{2}) \nonumber \\
&& +\frac{1}{8ac^{2}}[\dot{x}^{2}+\dot{y}^{2}+2(x\dot{y}-\dot{x}y)
+x^{2}+y^{2}] \nonumber \\
&&  \cdot[3(\dot{x}^{2}+\dot{y}^{2}) +2(x\dot{y}-\dot{x}y)-(x^{2}+y^{2})] \nonumber \\
&&  +\frac{3}{2ac^{2}}(\frac{1-\mu}{r_1}+\frac{\mu}{r_2}) \nonumber \\
&& \cdot[\dot{x}^{2}+\dot{y}^{2}-x^{2}-y^{2}-\mu(1-\mu)], \\
C_J &=& x^{2}+y^{2}+2(x\dot{y}-\dot{x}y)-\dot{x}^{2}-\dot{y}^{2}.
\end{eqnarray}
Notice that $C_J$ is the Jacobi constant when the Newtonian
problem $\mathcal{L}_0$ is considered only, but is not in the
present PN problem.

If Eq. (5) is considered, we have the acceleration equations
\begin{eqnarray}
\ddot{x} &\approx& \mathcal{L}_{x}+(1+\frac{\omega_1}{c^{2}})\dot{y} \nonumber \\
& & -\frac{1}{2ac^{2}}
[\dot{x}^{2}+\dot{y}^{2}+2(x\dot{y}-\dot{x}y)  +x^{2}+y^{2} \nonumber \\
& & +6(\frac{1-\mu}{r_1}+\frac{\mu}{r_2})](\ddot{x}_{0}-\dot{y})  \nonumber \\
&& -\frac{1}{ac^{2}}(\dot{x}-y)\{\ddot{x}_{0}(\dot{x}-y)+\ddot{y}_{0}(\dot{x}+y) \nonumber \\
& & +x\dot{x}+y\dot{y}-\frac{3}{r^{3}_{1}}(1-\mu)[y\dot{y}+(x+\mu)\dot{x}] \nonumber \\
&& -\frac{3\mu}{r^{3}_{2}}[y\dot{y}+(x-1+\mu)\dot{x}]\}, \\
\ddot{y} &\approx& \mathcal{L}_{y} -(1+\frac{\omega_1}{c^{2}})\dot{x}  \nonumber \\
& & -\frac{1}{2ac^{2}}
[\dot{x}^{2}+\dot{y}^{2}+2(x\dot{y}-\dot{x}y) +x^{2}+y^{2} \nonumber \\
& &  +6(\frac{1-\mu}{r_1}+\frac{\mu}{r_2})](\ddot{y}_{0}+\dot{x}) \nonumber \\
& & -\frac{1}{ac^{2}}(\dot{y}+x) \{\ddot{x}_{0}(\dot{x}-y)+\ddot{y}_{0}(\dot{x}+y)  \nonumber \\
& & +x\dot{x}+y\dot{y}-\frac{3}{r^{3}_{1}}(1-\mu) [y\dot{y}+(x+\mu)\dot{x}] \nonumber \\
& &  -\frac{3\mu}{r^{3}_{2}}[y\dot{y}+(x-1+\mu)\dot{x}]\}.
\end{eqnarray}
In the above equations, $\ddot{x}_{0}$ and $\ddot{y}_{0}$ denote
the accelerations from the Newtonian term $\mathcal{L}_0$. In
fact, $\ddot{x}_{0}$ and $\ddot{y}_{0}$ should be $\ddot{x}$ and
$\ddot{y}$, respectively. In this way, there are implicit
acceleration equations suggested in [12]. They are given
coherently, and should be completely equivalent to Eqs. (15)-(20).
When the accelerations are solved explicitly from the coherent
implicit acceleration equations, they have long and complicated
expressional forms. Fortunately, Eqs. (15)-(20) have simple
explicit expressional forms. It is clear that Eqs. (23) and (24)
are obtained via $\ddot{x}\rightarrow\ddot{x}_{0}$ and
$\ddot{y}\rightarrow\ddot{y}_{0}$ in the right functions of the
coherent implicit acceleration equations. In other words, the PN
terms of the accelerations $\ddot{x}$ and $\ddot{y}$ in the right
functions of the coherent implicit acceleration equations are
truncated. In this sense, Eqs. (23) and (24), as 1PN equations of
motion, are the incoherent equations of motion for the Lagrangian
$\mathcal{L}$ in Eq. (8). Since this system is conservative, the
energy should be a conserved quantity. However, we have no way to
give the energy an exact expression when the incoherent equations
of motion are used. In this sense, Eqs. (23) and (24)
approximately conserve the energy (21).

In short, the 1PN Lagrangian $\mathcal{L}$ in Eq. (8) has two
descriptions of the equations of motion, namely, the coherent
equations of motion (i.e. the differential equations of the
generalized momenta) and the incoherent ones (namely, the
acceleration equations or the differential equations of the
velocities). What results are caused by the two distinct
treatments to the equations of motion? Its answer awaits  detailed
numerical comparisons.

\subsection{Solar System}

First, we consider only the Newtonian circular restricted
three-body problem $\mathcal{L}_0$ in Eq. (9) in the Solar System.
An eighth- and ninth-order Runge-Kutta-Fehlberg explicit
integration algorithm [RKF8(9)] of a variable time step is chosen
as a numerical integrator. We take the dimensionless mass
parameter $\mu= 0.001$, which approaches to  the ratio of Jupiter
to the Sun. The Jacobi constant in Eq. (22) is given by
$C_J=3.07$. Two orbits with the initial values of $x$=0.55 and
0.56 are chosen. They have the same initial values $y=\dot{x}=0$,
but their initial values $\dot{y}$ $(>0)$ can be solved from Eq.
(22) and are different. It is shown via the Poincar\'{e} surface
of section in Fig. 1(a) that orbit 1 with the initial value
$x=0.55$ forms three islands and therefore is regular. It is clear
that orbit 2 with the initial value $x=0.56$ has  many random
points in a large region and thus exhibits a chaotic behavior.
Although RKF8(9) is not an energy-preserving method, it gives such
a high accuracy in the magnitude of an order $10^{-13}$ to the
energies of the two orbits in Fig. 1(b).

Then, we focus on the PN circular restricted three-body problem
$\mathcal{L}$ in Eq. (8) in the Solar System. As claimed in [14],
$a=1$ and $c=10^{4}$ are recommended. $1/c^{2}=10^{-8}$ is near to
the 1PN relativistic effect of the Sun and Jupiter. The PN terms
$\mathcal{L}_1$ and $\mathcal{L}_2$ are so small that they have
negligible effects on the structure of phase space of orbits 1 and
2 in the Newtonian problem when the incoherent Eqs. (23) and (24)
are used. Because of this, the accuracy of energies of the two
orbits in the present case is almost the same as that in the
Newtonian problem although the energy $E$ in Eq. (21) is
approximately conserved by Eqs. (23) and (24). These results are
not typically different when the incoherent Eqs. (23) and (24)
give place to the coherent Eqs. (15)-(20). That is to say, for
such weak relativistic effects in the Solar System, the incoherent
equations and the coherent ones provide almost the same accuracy
of energy. As a point to illustrate, the conservation of the
Jacobi integral in the PN circular restricted three-body problem
[14] is still based on the use of the incoherent equations. In
addition, there is no explicit difference in the structure of
phase space of orbits on the Poincar\'{e} surface of section
between the two cases. Namely, an orbit is not chaotic in the
coherent equations if it is ordered in the incoherent ones;
inversely, an orbit is not regular in the incoherent equations if
it is chaotic in the coherent ones.

Although the orbital dynamical features of order and chaos are
absolutely dominated by the Newtonian term $\mathcal{L}_0$ and are
not typically affected by the weak relativistic effects, the
incoherent equations and the coherent ones still lead to somewhat
differences in the positions and velocities of orbits. Even there
are large differences in some cases. For the ordered orbit 1 in
Fig. 2, the separation between the positions in the two cases is
small and grows nearly linearly with time. However, this
separation becomes typically large and increases exponentially for
the chaotic orbit 2. These results should be reasonable. The
difference between the incoherent equations and the coherent ones
is the 2PN relativistic effect having an order of $10^{-16}$. The
chaoticity of orbit 2 makes the difference larger and larger with
the integration time increasing. As a noticeable point, the
separation remains invariant at the value 1 after $t=1168$. This
is due to the saturation of chaotic orbits in a bounded region.
This problem also occurs in computations of Lyapunov exponents of
two nearby orbits [15, 16]. The renormalization must be
considered. As a result, the saturation of orbits is avoided and
the Lyapunov exponents of two nearby orbits can be computed in a
long enough time.

\subsection{Compact objects}

Now, assume the three bodies consisting of compact objects, such
as neutron stars and/or black holes. Here, $c=1$ is given as in
the usual relativistic issue. In this case, $1/a$ in the PN parts
$\mathcal{L}_1$ and $\mathcal{L}_2$ plays book-keeping for the 1PN
relativistic effect and thus $a\gg 1$ is required. Setting
$a=1345$, the 1PN relativistic effect in an order of $10^{-3}$ in
the case of compact objects is much stronger than that in an order
of $10^{-8}$ in the Solar System. We take only orbit 1 as an
example and compare the related differences between the incoherent
equations and the coherent ones.

It can be seen clearly from Fig. 3 (a) and (b) that orbit 1 on the
Poincar\'{e} surface of section becomes chaotic in the incoherent
equations, whereas it is still regular in the coherent equations.
Therefore, it is not unexpected that the difference between the
incoherent equations and the coherent ones may lead to the two
sets of equations having different dynamical behaviors of order
and chaos. The energy accuracy with an order of $10^{-5}$ is poor
in the incoherent equations in Fig. 3(c). If the distance between
the binaries gets smaller and smaller, e.g. $a=100, 50,
10,\cdots$, then the energy accuracy becomes poorer and poorer.
Particularly for $a=1$, the energy cannot be conserved numerically
at all, as claimed in [14]. These results completely satisfy the
theoretical need because the energy is not an integral of motion
that can be conserved exactly by the incoherent equations. On the
contrary, the energy accuracy is still high enough and reaches an
order of $10^{-13}$ in the coherent equations in Fig. 3(d). The
extremely high accuracy of energy does not depend on the value of
$a$.  This is owing to the energy as an exact integral of the
coherent equations.

\section{PN Lagrangian systems of compact binaries}

In a spinless compact binary system, the binaries have masses
$M_1$ and $M_2$. We take the total mass $M=M_1+M_2$, the reduced
mass $\mu=M_1M_2/M$, the mass ratio $\beta=M_2/M_1$ and the
dimensionless mass parameter $\eta=\mu/M=\beta(1+\beta)^{-2}$. The
position and velocity of body 1 relative to body 2 are
$\mathbf{r}=(x,y,z)$ and $\mathbf{v}=(\dot{x},\dot{y},\dot{z})$,
respectively. $r=|\mathbf{r}|$ represents the relative distance of
the binaries, and $v=|\mathbf{v}|$ is the magnitude of the
relative velocity. Scale transformations are given to the
coordinate $\mathbf{r}$ and time $t$ as follows:
$\mathbf{r}\rightarrow M \mathbf{r}$ and $t\rightarrow Mt$.

According to two cases, the dynamical differences off the binaries
between the incoherent equations of motion and the coherent ones
are compared.

\subsection{Two bodies nonspinning}

The evolution of the binaries is described by the dimensionless PN
Lagrangian formulation
\begin{equation}
\ell=\ell_n+\ell_{1pn},
\end{equation}
where the first term is the Newtonian two-body problem
\begin{equation}
\ell_n=\frac{v^2}{2}+\frac{1}{r},
\end{equation}
and the second term is the 1PN contribution [17]
\begin{eqnarray}
\ell_{1pn} &=&
\frac{1}{c^{2}}\{\frac{1-3\eta}{8}v^4+\frac{1}{2r}[\frac{\eta}{r^2}
(\mathbf{r}\cdot\mathbf{v})^2 \nonumber \\
&&  +(3+\eta)v^2-\frac{1}{r}]\}.
\end{eqnarray}

The generalized momenta satisfy the following algebraic equation
\begin{eqnarray}
\mathbf{p} &=& \mathbf{v}+\frac{1}{c^{2}}
\{\frac{v^2}{2}(1-3\eta)\mathbf{v} +\frac{1}{r}[\frac{\eta}{r^2}
(\mathbf{r}\cdot\mathbf{v})\mathbf{r} \nonumber \\
& &  +(3+\eta)\mathbf{v}]\}.
\end{eqnarray}
As was mentioned in Sect. 2, the Lagrangian $\ell$ has a coherent
1PN generalized momentum differential equation
\begin{eqnarray}
\frac{d\mathbf{p}}{dt} &=&
-\frac{\mathbf{r}}{r^3}\{1+\frac{1}{c^{2}}[\frac{3\eta}{2r^2}(\mathbf{r}\cdot\mathbf{v})^2
+\frac{3+\eta}{2}v^2-\frac{1}{r}]\}  \nonumber \\
&&  +\frac{\eta}{c^2r^3}(\mathbf{r}\cdot\mathbf{v})\mathbf{v}.
\end{eqnarray}
Eq. (3) is also included in the coherent equations of motion.
Using Eq. (28), we have the velocity
\begin{eqnarray}
\mathbf{v} &=&
[\mathbf{p}-\frac{\eta}{c^2r^3}(\mathbf{r}\cdot\mathbf{v})\mathbf{r}]
/[1+\frac{1}{c^{2}}(\frac{1-3\eta}{2} v^2 \nonumber \\
& & +\frac{3+\eta}{r})].
\end{eqnarray}
Because $1-3\eta=(\beta^2-\beta+1)/(\beta+1)^{2}>0$, the
denominator in Eq. (30) is always larger than 1. This shows that
the iterative solution of $\mathbf{v}$  is convergent. Eqs. (3),
(29) and (30) exactly conserve  the energy
\begin{eqnarray}
\mathcal{E} &=& \frac{v^2}{2}-\frac{1}{r}+\frac{1}{c^{2}}
\{\frac{3}{8}(1-3\eta)v^4+\frac{1}{2r}[(3+\eta)v^2  \nonumber \\
&& +\frac{\eta}{r^2}(\mathbf{r}\cdot\mathbf{v})^2 +\frac{1}{r}]\}.
\end{eqnarray}
They also exactly conserve the orbital angular momentum vector
\begin{equation}
\mathbf{L}=\mathbf{r}\times\mathbf{p}=[1+\frac{1}{c^{2}}(\frac{1-3\eta}{2}
v^2+\frac{3+\eta}{r})]\mathbf{r}\times\mathbf{v}.
\end{equation}

The 1PN relative acceleration is written as
\begin{equation}
\mathbf{\ddot{r}}=\mathbf{a}_n+\mathbf{a}_{1pn},
\end{equation}
where the two parts are
\begin{equation}
\mathbf{a}_n=-\frac{\mathbf{r}}{r^3},
\end{equation}
\begin{eqnarray}
\mathbf{a}_{1pn} &=& -\frac{1}{r^{2}c^{2}}
\{\frac{\mathbf{r}}{r}[(1+3\eta)v^2
 \nonumber \\
&& -\frac{2}{r}(2+\eta) -\frac{3\eta}{2r^2}(\mathbf{r}\cdot\mathbf{v})^2]  \nonumber \\
&&  -\frac{2}{r}(2-\eta)(\mathbf{r}\cdot\mathbf{v})\mathbf{v}\}.
\end{eqnarray}
Eq. (33) approximately conserves the energy (31) or the 1PN
Hamiltonian quantity
\begin{eqnarray}
\mathcal{H}(\mathbf{r},\mathbf{p}) &=& \mathbf{v}\cdot
\mathbf{p}-\ell \approx\frac{p^2}{2}-\frac{1}{r}+\frac{1}{c^{2}}
\{\frac{3\eta-1}{8}p^4  \nonumber \\
&&
-\frac{1}{2r}[(3+\eta)p^2+\frac{\eta}{r^2}(\mathbf{r}\cdot\mathbf{p})^2
-\frac{1}{r}]\}.
\end{eqnarray}
The symbol $``\approx"$ means that $\mathcal{E}$ is not exactly
equal to $\mathcal{H}$ because the higher-order (2PN, 3PN,
$\cdots$) terms are truncated in $\mathcal{H}$ but such
truncations do not occur in $\mathcal{E}$. In addition, Eq. (33)
approximately conserves the orbital angular momentum vector (32).

\begin{table*}[tbp]
\caption{In the nonspinning compact binary system, the separation
between the solution $(x_{ce},y_{ce})$ in the coherent equations
and the solution $(x_{ie},y_{ie})$ in the incoherent ones,
$D=\sqrt{(x_{ce}-x_{ie})^2+(y_{ce}-y_{ie})^2}$.\label{tab1}}
 \begin{tabular}{cccccccc}
  \hline\hline
   $t$ & $x_{ie}$ & $x_{ce}$ & $y_{ie}$ &  $y_{ce}$ & $D$ \\
 \hline
    1 & 17.03913 & 17.03910 & 10.09351 &  10.09349 & 0.000036 \\
    10 & 16.9548 & 16.95137 & 10.89020 &  10.88785  & 0.0042 \\
    100 & 9.39192 & 9.03218 & 13.34784 &  13.03775  & 0.47 \\
   1000 & 2.97387 & 0.67666 & -2.62049 &  7.82169  & 10.7 \\
   10000 & 15.05357 & -14.83949 & -4.46476 & 8.48086  & 32.6 \\
\hline \hline
\end{tabular}
\end{table*}

Setting the speed of light $c=1$, the mass ratio $\beta=5/4$ and
the initial conditions ($x=17.04$, $y=10$, $\dot{y}=0.094$,
$z=\dot{x}=\dot{z}=0$), we apply RKF8(9) to numerically integrate
the coherent Eqs. (3) and (29) or the incoherent Eq. (33). In Fig.
4 (a) and (b), the energy error is very large in the magnitude of
an order of 0.1  in the incoherent equations, but it is so small
that it arrives at an order of $10^{-12}$ in the coherent
equations when the integration time $t=10^{6}$. In Fig. 4 (c) and
(d), the errors of the orbital angular momentum in the two sets of
equations of motion  are also similar to those of the energy.
These results strongly support the preference of the coherent
equations over the incoherent equations in the conservation of the
energy and the angular momentum.

It should be emphasized that the chosen orbit is limited to the
plane $z=0$ due to the conserved angular momentum  (32) with the
initial angular momentum $\mathbf{L}=(0,0,L_z)$. Additionally, the
system $\ell$ is integrable and regular because its formal
equivalent Hamiltonian (that is not $\mathcal{H}$ and is not
easily written in detail) has four independent integrals of motion
in the six-dimensional phase space [9]. That means that any orbit
in the system $\ell$  is always nonchaotic regardless of the
choice of the incoherent equations and the coherent ones. In spite
of this, the separation between the solution of the coherent
equations and that of the incoherent equations will become rather
large as the integration time is long enough in Table 1.

\subsection{Two bodies spinning}

Now, let us consider that the binaries in Eq. (25) are spinning,
and their spin effects are restricted to spin-orbit coupling
interaction with a 1.5 PN accuracy
\begin{equation}
\ell_{so}=\frac{\eta}{c^{3}r^{3}}\mathbf{v}\cdot [\mathbf{r}\times
(\gamma_1\mathbf{S}_1+\gamma_2\mathbf{S}_2)],
\end{equation}
where $\gamma_1 = 2+3\beta/2$, $\gamma_2 = 2+3/(2\beta)$, and
$\mathbf{S}_1$, $\mathbf{S}_2$ are measured in terms of $\mu M$.
The spin magnitudes are $S_1=|\mathbf{S}_1|=\chi_1M^{2}_{1}/(\mu
M)$ and $S_2=|\mathbf{S}_2|=\chi_2M^{2}_{2}/(\mu M)$, where
$0\leq\chi_1\leq1$ and $0\leq\chi_2\leq1$ are dimensionless spin
parameters. Under the Newton-Wigner-Pryce spin supplementary
condition [18], the spin-orbit coupling term does not depend on
accelerations. When the spin-orbit term is included in Eq. (25),
the PN Lagrangian system becomes
\begin{equation}
\mathbb{L}=\ell+\ell_{so}.
\end{equation}

The generalized momentum of $\mathbb{L}$ is
\begin{equation}
\mathbf{P}=\mathbf{p}+\frac{\eta}{c^{3}r^{3}} \mathbf{r}\times
(\gamma_1\mathbf{S}_1+\gamma_2\mathbf{S}_2).
\end{equation}
The PN Lagrangian $\mathbb{L}$ has a coherent 1.5PN differential
equation of the generalized momentum
\begin{eqnarray}
\frac{d\mathbf{P}}{dt} &=& \frac{\partial\ell}{\partial
\mathbf{r}}+\frac{3\eta\mathbf{r}}{c^3r^{5}}(\mathbf{r}\times\mathbf{v})
\cdot(\gamma_1\mathbf{S}_1+\gamma_2\mathbf{S}_2) \nonumber \\
& & -\frac{\eta}{c^3r^{3}}\mathbf{v}\times
(\gamma_1\mathbf{S}_1+\gamma_2\mathbf{S}_2).
\end{eqnarray}
The spin precession equations are given by
\begin{eqnarray}
\mathbf{\dot{S}}_1 &=& \mathbf{S}_1\times\frac{\partial
\mathbb{L}}{\partial \mathbf{S}_1}
=\frac{\eta\gamma_1}{c^{3}r^{3}}(\mathbf{r}\times \mathbf{v})\times \mathbf{S}_1, \\
\mathbf{\dot{S}}_2 &=& \mathbf{S}_2\times\frac{\partial
\mathbb{L}}{\partial \mathbf{S}_2}
=\frac{\eta\gamma_2}{c^{3}r^{3}}(\mathbf{r}\times
\mathbf{v})\times \mathbf{S}_2.
\end{eqnarray}
The velocity has an iterative form
\begin{eqnarray}
\mathbf{v} &=& [\mathbf{P}-\frac{\eta}{c^2r^3}(\mathbf{r}\cdot\mathbf{v})\mathbf{r}\nonumber \\
&&  -\frac{\eta}{c^3r^{3}}\mathbf{r}\times
(\gamma_1\mathbf{S}_1+\gamma_2\mathbf{S}_2)] \nonumber \\
&& /[1+\frac{1}{c^{2}}(\frac{1-3\eta}{2}v^2+\frac{3+\eta}{r})].
\end{eqnarray}
Eqs. (3), (40)-(43) are the coherent equations of motion with
respect to the PN Lagrangian $\mathbb{L}$. The energy (31) is
exactly conserved by these equations of motion. There is the
conserved total angular momentum vector
\begin{eqnarray}
\mathbf{J} &=& \mathbf{r}\times \mathbf{P}+\frac{1}{c}(\mathbf{S}_1+\mathbf{S}_2)  \nonumber \\
&=& \mathbf{L}+\frac{\eta}{c^{3}r^{3}}\mathbf{r}\times
[\mathbf{r}\times (\gamma_1\mathbf{S}_1 +\gamma_2\mathbf{S}_2)]  \nonumber \\
&& +\frac{1}{c}(\mathbf{S}_1+\mathbf{S}_2).
\end{eqnarray}
The spin magnitudes $S_1$ and  $S_2$ are exact constants of
motion, too.

The 1.5PN acceleration equation is expressed as
\begin{equation}
\mathbf{a}=\mathbf{a}_n+\mathbf{a}_{1pn}+\mathbf{a}_{so},
\end{equation}
where the third term is
\begin{eqnarray}
\mathbf{a}_{so} &=&
\frac{\eta}{c^{3}r^{3}}[\frac{3\mathbf{r}}{r^{2}}
(\mathbf{r}\times\mathbf{v})\cdot (\gamma_1\mathbf{S}_1+\gamma_2\mathbf{S}_2) \nonumber \\
&& -2 \mathbf{v}\times(\gamma_1\mathbf{S}_1 +\gamma_2\mathbf{S}_2) \nonumber \\
&& +\frac{3}{r^{2}}(\mathbf{r}\cdot\mathbf{v})
\mathbf{r}\times(\gamma_1\mathbf{S}_1+\gamma_2\mathbf{S}_2)].
\end{eqnarray}
Eq. (45) is approximately provided because of the higher-order
terms truncated. Eqs. (41), (42) and (45) approximately conserve
the energy (31) or the 1.5PN Hamiltonian
\begin{equation}
\mathbb{H} =
\mathcal{H}(\mathbf{r},\mathbf{P})+\frac{\eta}{c^{3}r^{3}}\mathbf{r}\cdot
[\mathbf{P}\times (\gamma_1\mathbf{S}_1+\gamma_2\mathbf{S}_2)].
\end{equation}
They also approximately conserve the total angular momentum vector
(44), but exactly conserve the spin magnitudes $S_1$ and  $S_2$.

Given $\chi_1=\chi_2=1$, the initial spin vectors are chosen as
$\mathbf{S}_1=S_1\mathbf{\hat{S}}_1$ and
$\mathbf{S}_2=S_2\mathbf{\hat{S}}_2$, where $\mathbf{\hat{S}}_1$
and $\mathbf{\hat{S}}_2$ are two unit vectors
\begin{eqnarray}
\mathbf{\hat{S}}_1 &=& (0.1,0.3,0.5)/\sqrt{0.1^2+0.3^2+0.5^2}, \nonumber \\
\mathbf{\hat{S}}_2 &=& (0.7,0.3,0.1)/\sqrt{0.7^2+0.3^2+0.1^2}.
\nonumber
\end{eqnarray}
The mass ratio and the orbit in Sect. 3 are still used. The
accuracies of the energy and the total angular momentum in the
coherent equations of motion are much better than those in the
incoherent ones for the two bodies spinning, as shown in Fig. 5.
This result is very similar to that for the two bodies
nonspinning. However, unlike in the case of the two bodies
nonspinning, the spin-orbit effects lead to the orbit precessing
in the direction $z$. Fig. 6 clearly shows that the precession
values of $z$ in the coherent equations are very inconsistent with
those in the incoherent ones after a short time.

To know whether the solutions between the coherent equations and
the incoherent ones have the same chaotic behavior, we rely on the
largest Lyapunov exponent of two nearby orbits [15]
\begin{equation}
\lambda=\lim_{t\rightarrow\infty}\frac{1}{t}\ln\frac{d(t)}{d(0)},
\end{equation}
where $d(t)$ and $d(0)$ denote the separations between the two
nearby orbits at times $t$ and 0. The Lyapunov exponent of the
orbit in the incoherent equations tends to zero in Fig. 7, and
therefore the solutions of the incoherent equations are regular.
However, the bounded solutions of the coherent equations should be
chaotic due to the presence of a positive Lyapunov exponent. It is
clear that the time for the Lyapunov exponent tending to zero or a
stabilizing positive value is long enough, e.g. $t=10^{7}$. By
contrast, a fast Lyapunov indicator can distinguish between the
two cases of order and chaos with less computational cost. The
fast Lyapunov indicator of two nearby orbits [16] is calculated by
\begin{equation}
\Lambda=\log_{10}\frac{d(t)}{d(0)}.
\end{equation}
This indicator $\Lambda$ increasing in a power law with time
$\log_{10} t$ indicates  the regularity of the solutions in the
incoherent equations, whereas the indicator having an exponential
growth with time shows the chaoticity of the solutions in the
coherent equations. Only when the time arrives at $3\times10^{4}$
in Fig. 8, can the two cases of order and chaos  clearly be
distinguished.

In a word, the methods of Lyapunov exponents and fast Lyapunov
indicators have confirmed together that the orbit is ordered in
the incoherent equations but chaotic in the coherent ones. Of
course, it is possible that an orbit  is chaotic in the incoherent
equations but regular in the coherent ones when the dynamical
parameters, the initial conditions and the initial spin vectors
are altered. These results are due to the nonintegrability of the
PN Lagrangian $\mathbb{L}$ and the nonequivalence between the two
sets of equations. On one hand, the Lagrangian $\mathbb{L}$ has
its formal exact equivalent Hamiltonian  (nonequal to
$\mathbb{H}$) [9], which includes higher-order spin-spin coupling
terms and contains four independent integrals of the energy (31)
and the total angular momentum vector (44). When the canonical,
conjugate spin variables of [19] are adopted, this Hamiltonian has
a ten-dimensional phase space and therefore is nonintegrable.
Namely, the Lagrangian $\mathbb{L}$ is nonintegrable and may be
chaotic.  By contrast, the 1.5PN Hamiltonian $\mathbb{H}$ is
integrable and nonchaotic because of the existence of a fifth
integral, the length of the orbital angular momentum vector
$\mathbf{r}\times\mathbf{P}$ [19]. On the other hand, the
incoherent equations and the coherent ones exist the difference of
2PN order terms and are only approximately related two different
systems. It is supported in dynamical systems theory that one
system is chaotic but an approximately related system can be
ordered. Therefore, it should be reasonable that the solutions of
the incoherent equations and those of the coherent equations have
different dynamical behaviors under a certain circumstance.

It is worth pointing out that the difference between the
incoherent equations and the coherent ones  are too small to
affect the solutions of the two sets of equations for regular
orbits in the PN circular restricted three-body problem in the
Solar System. In other words, the two sets of equations are almost
equivalent and have the same dynamical behavior. This result is
also suitable for spinless binaries or spinning bodies in the
Solar System. Thus, the famous PN effects like perihelion or
periastron advances for spinless binaries or the geodetic, the
Lense-Thirring and the Schiff precessions for spinning bodies are
given the same results by the two sets of equations in the Solar
System.

\section{Summary}

Based on the construction of PN Lagrangian equations of motion in
the above examples, several points can be concluded as follows.

There are two paths to obtain the equations of motion from a PN
Lagrangian formalism. As one path, the total accelerations at the
same PN order of the Lagrangian are derived from the
Euler-Lagrangian equations of this Lagrangian, by truncating
higher-order terms of the accelerations in the Euler-Lagrangian
equations. They are the incoherent PN equations of motion of the
Lagrangian system. Consequently, the constants of motion such as
the energy integral are only approximately conserved in the
incoherent equations. As another path, the differential equations
with respect to the generalized momenta directly come from the
Euler-Lagrangian equations, and also remain at the same PN order
of the Lagrangian. Although the velocities are not integration
variables, they can be solved from the generalized momentum
algebraic equations with an iterative method. Such equations of
motion exactly conserve the constants of motion in the PN
Lagrangian formalism, as Hamilton's equations exactly do in a PN
Hamiltonian formulation. In this way, the PN Lagrangian equations
of motion are coherent.

In fact, the incoherent equations of motion and the coherent ones
have somewhat differences and belong to approximately related two
different dynamical systems. When the differences associated to
the relativistic effects are such small quantities in the Solar
System, the accuracies of the constants of motion in the
incoherent equations of motion are almost the same as those in the
coherent ones if the adopted numerical integrator can achieve at a
high enough precision. The solutions of the incoherent equations
and those of the coherent equations have the same dynamical
behaviors of order and chaos. In addition, the differences between
the two sets of solutions are also small for the ordered case, but
large in a long time for the chaotic case. Unlike in the Solar
System, the differences between the two sets of equations of
motion are somewhat large in the strong gravitational systems of
compact objects. Because of this, the accuracies of the constants
of motion are very poorer in the incoherent equations of motion
even if the chosen integrator has a high precision, but very
better in the coherent equations of motion. The dynamical
behaviors may be different in the two cases. It is possible that
an orbit is chaotic in the incoherent equations while regular in
the coherent ones. It is also possible that an orbit  is ordered
in the incoherent equations whereas chaotic in the coherent ones.
These results are supported in dynamical systems theory. Even for
the regular case, the differences between the solutions in the two
sets of equations get larger and larger with time increasing.

There are no truncations when the coherent equations of motion are
derived from some PN Lagrangian problems. Therefore, the coherent
equations are worth recommending in the study of the dynamics and
gravitational waveforms of these PN Lagrangian systems.


\section*{Acknowledgments}

This research has been supported by the National Natural Science
Foundation of China under Grant No. 11533004 and the Natural
Science Foundation Innovation Group of Guangxi under Grant No.
2018GXNSFGA281007.



\newpage

\begin{figure*}[tbp]
\center{
\includegraphics[scale=0.65]{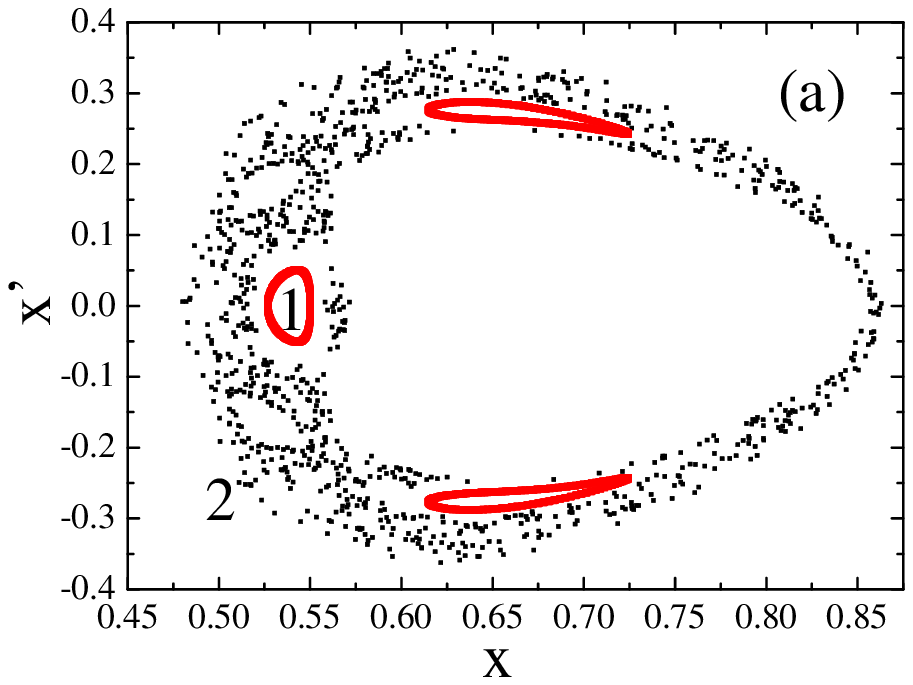}
\includegraphics[scale=0.65]{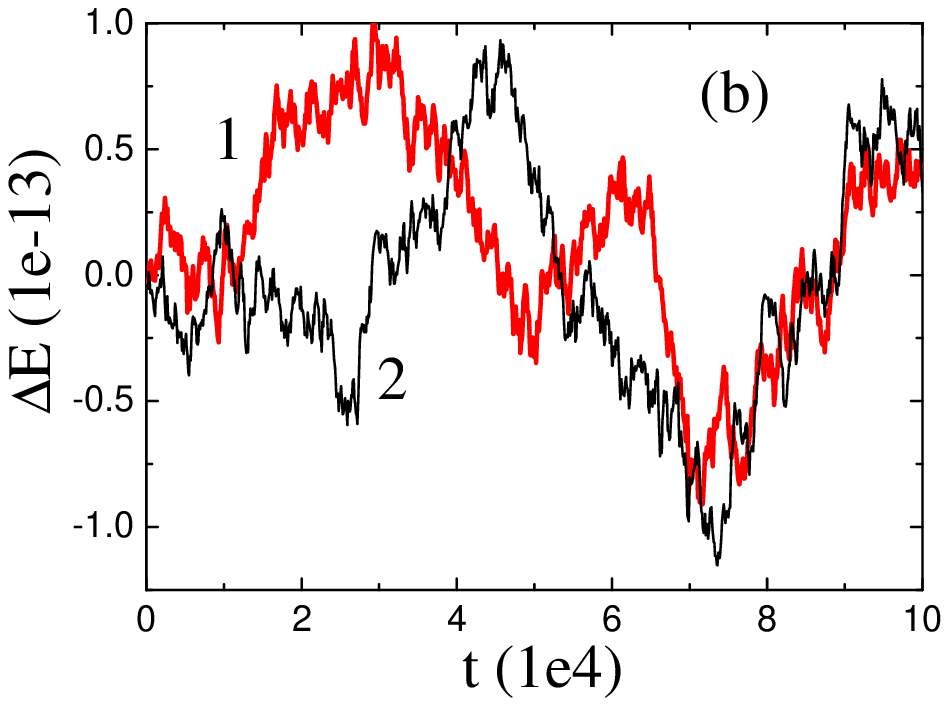}
\caption{The Newtonian circular restricted three-body problem: (a)
Poincar\'{e} surface of section $y=0$ and $\dot{y}>0$ and (b) the
energy error $\Delta E=E_0-E$, where $E_0$ is the initial energy.
Orbit 1 with the initial value $x=0.55$ is regular, whereas orbit
2 with the initial value $x=0.56$ is chaotic. }} \label{fig1}
\end{figure*}

\begin{figure*}[tbp]
\center{
\includegraphics[scale=0.7]{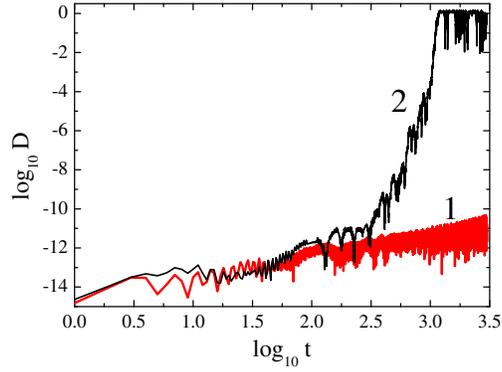}
\caption{The separations $D$ between the solution
$(x_{ie},y_{ie})$ in the incoherent equations and the solution
$(x_{ce},y_{ce})$ in the coherent ones,
$D=\sqrt{(x_{ie}-x_{ce})^{2}+(y_{ie}-y_{ce})^{2}}$. Orbits 1 and 2
are considered in the PN circular restricted three-body problem in
the Solar System. The separation of orbit 2 reaching 1 does not
grow after $t=1168$ due to the saturation of this bounded chaotic
orbit. }} \label{fig2}
\end{figure*}

\begin{figure*}[tbp]
\center{
\includegraphics[scale=0.65]{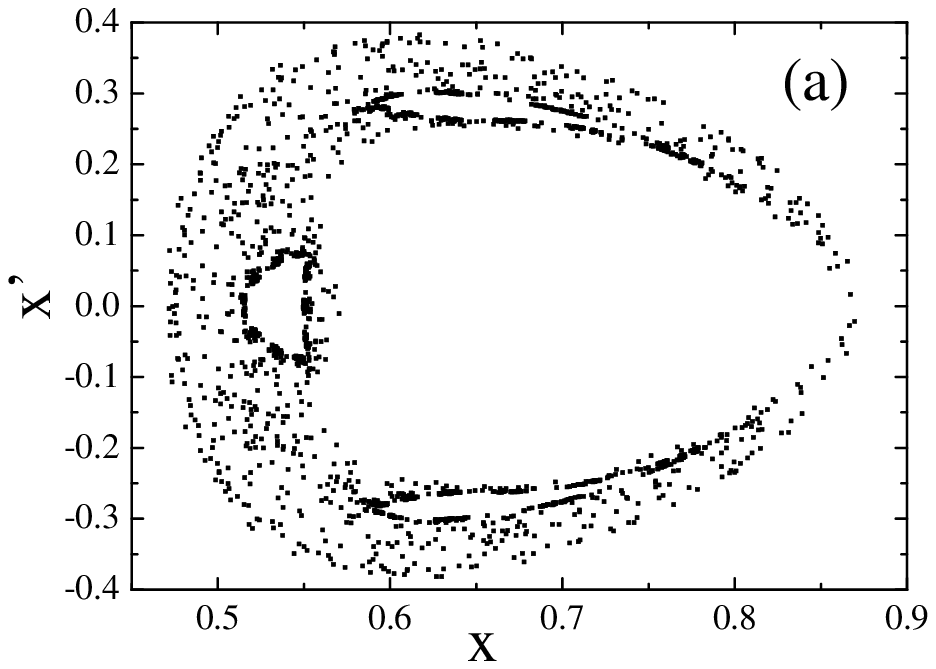}
\includegraphics[scale=0.65]{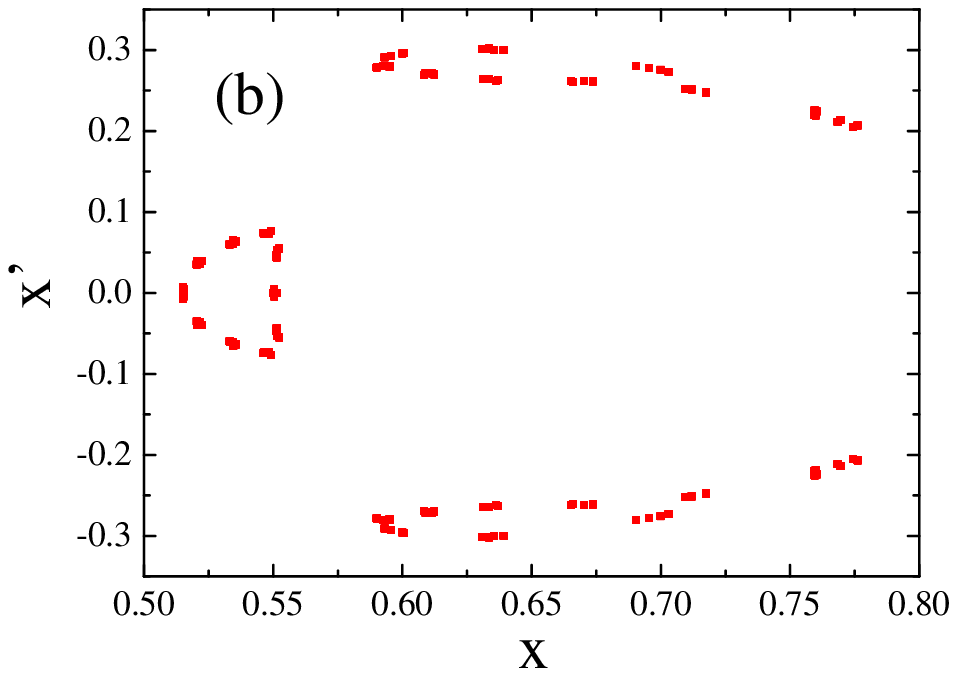}
\includegraphics[scale=0.65]{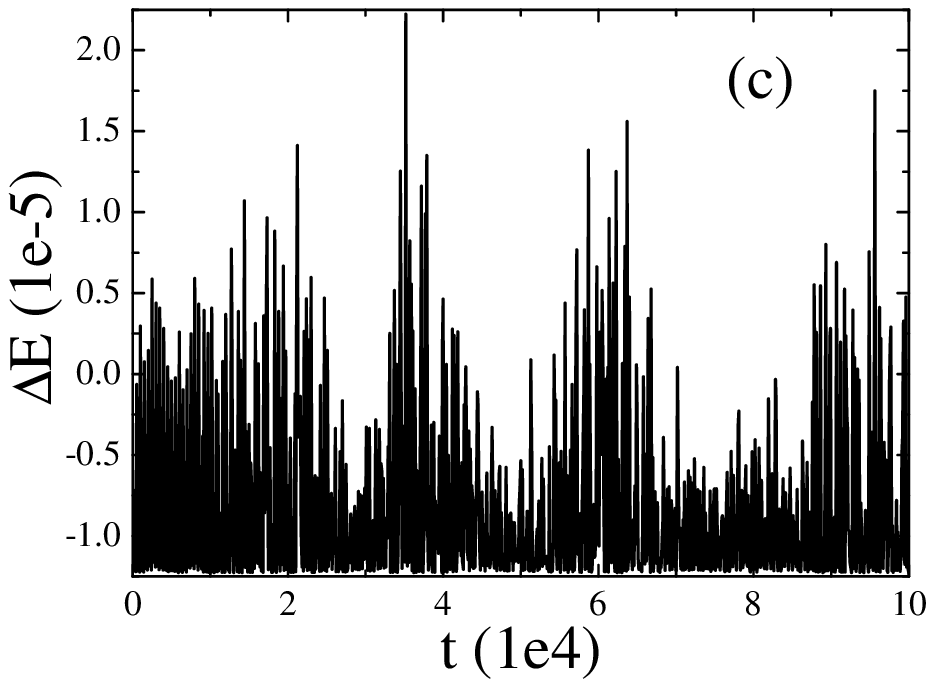}
\includegraphics[scale=0.65]{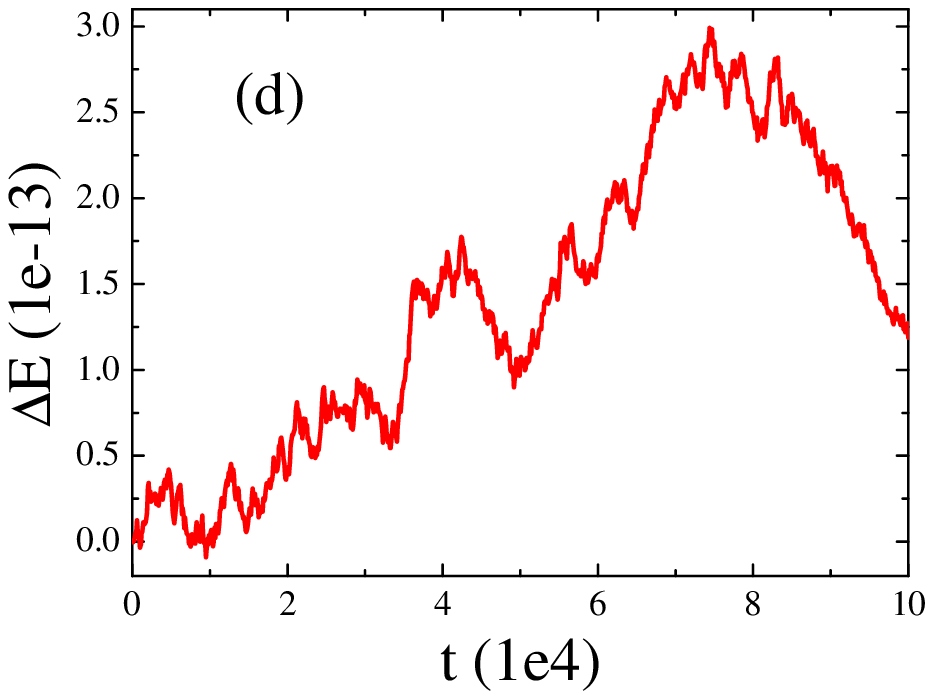}
\caption{The PN circular restricted three-body problem of compact
stars with the distance between the two primaries, $a=1345$: the
Poincar\'{e} surfaces of section of orbit 1 in the incoherent
equations of motion (a) and in the coherent ones (b); the energy
errors $\Delta E$ in the incoherent equations of motion (c) and in
the coherent ones (d). }} \label{fig3}
\end{figure*}

\begin{figure*}[tbp]
\center{
\includegraphics[scale=0.65]{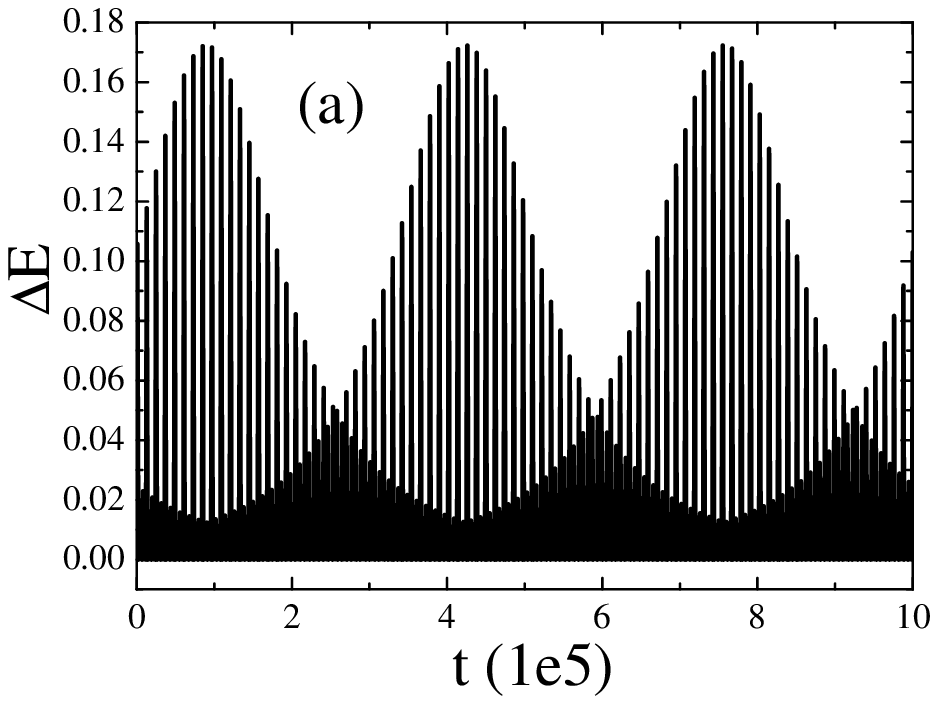}
\includegraphics[scale=0.65]{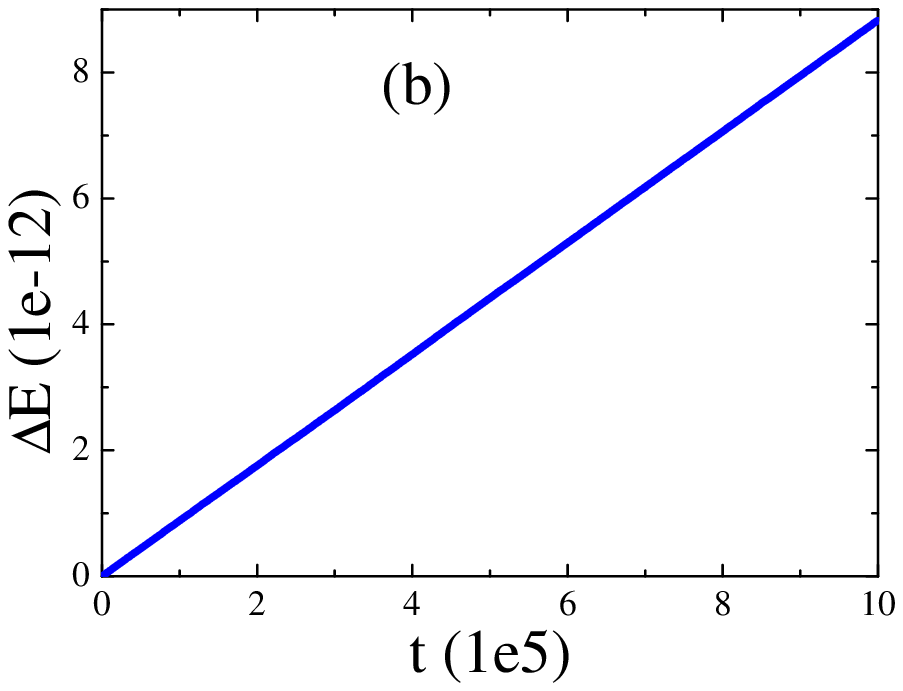}
\includegraphics[scale=0.65]{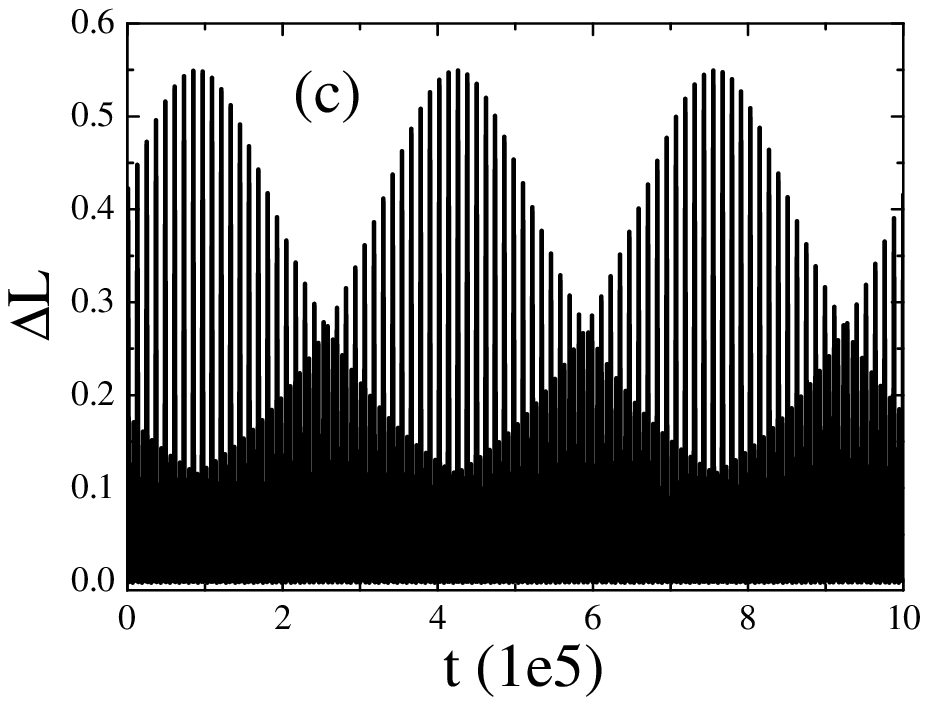}
\includegraphics[scale=0.65]{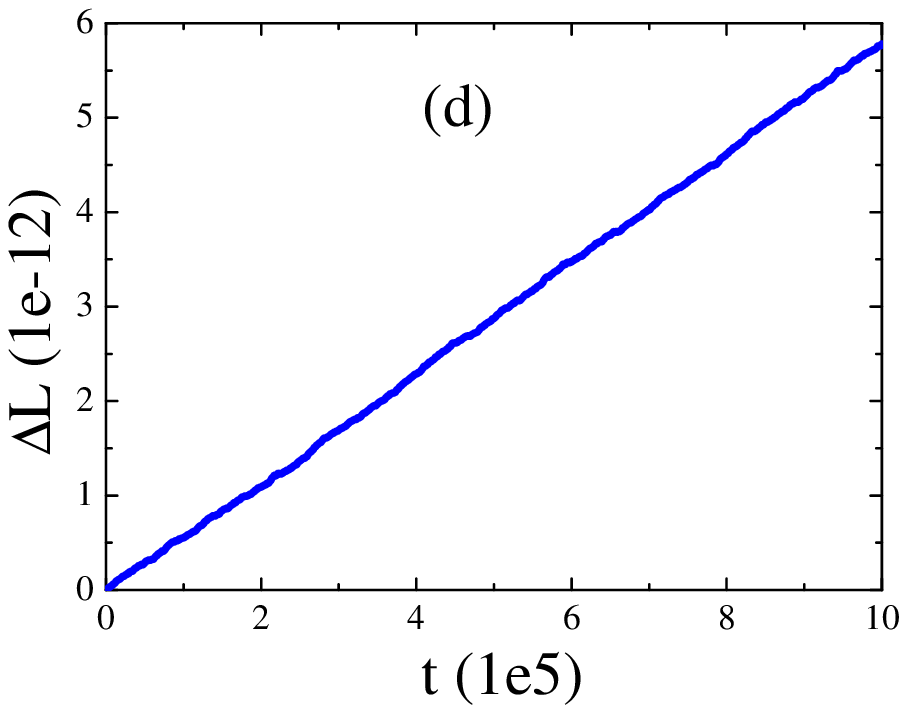}
\caption{The PN Lagrangian formulation of nonspinning compact
binaries: the energy errors $\Delta E$ in the incoherent equations
of motion (a) and in the coherent ones (b); the errors of the
angular momentum $\Delta L=L_0-L$ with $L=|\mathbf{L}|$ in the
incoherent equations of motion (c) and in the coherent ones (d).
}} \label{fig4}
\end{figure*}

\begin{figure*}[tbp]
\center{
\includegraphics[scale=0.65]{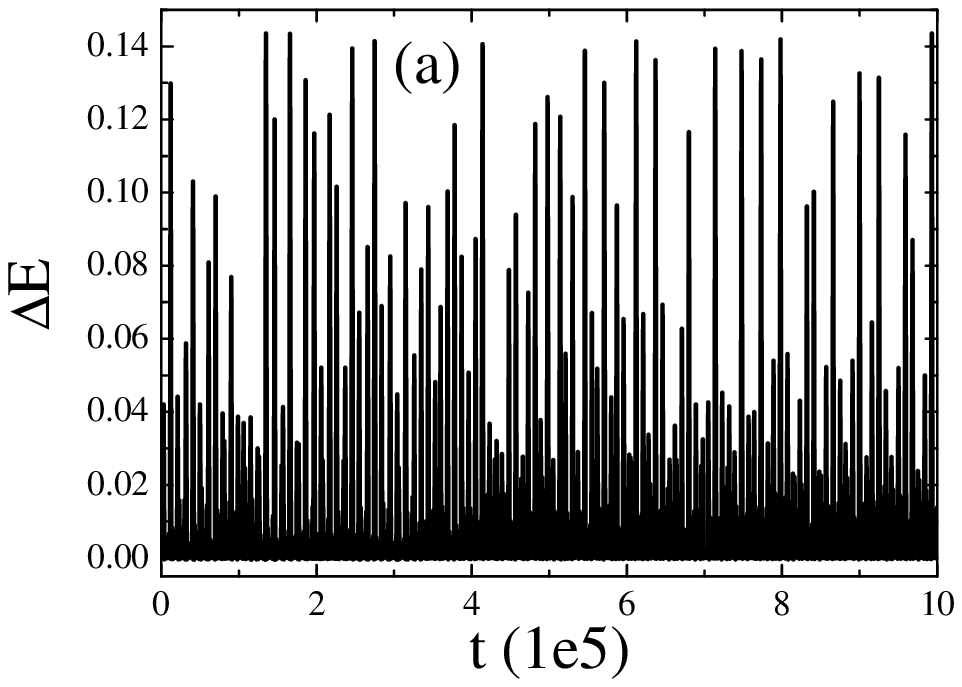}
\includegraphics[scale=0.65]{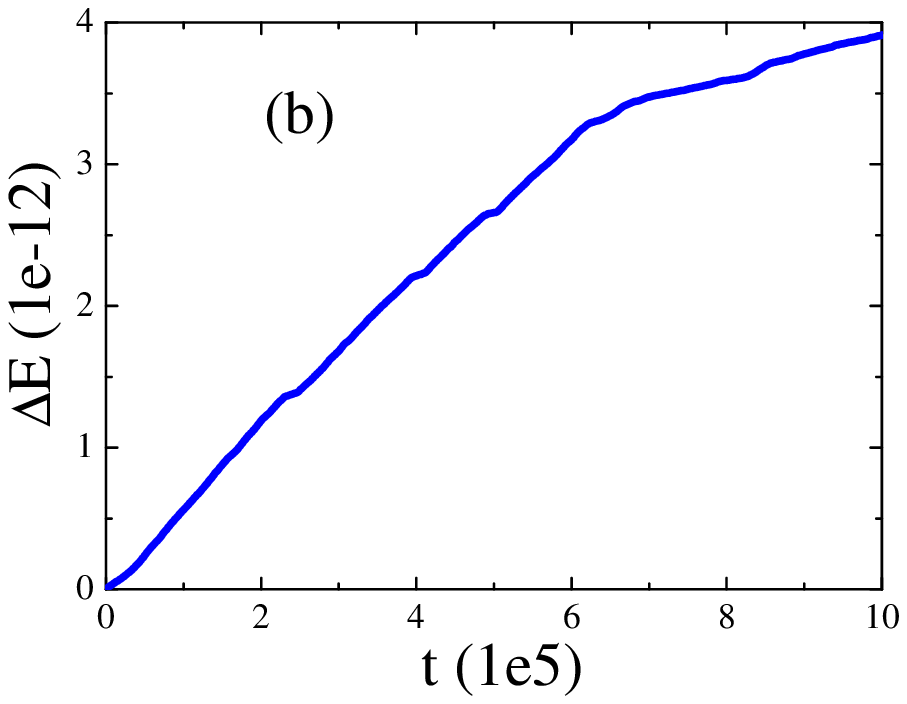}
\includegraphics[scale=0.65]{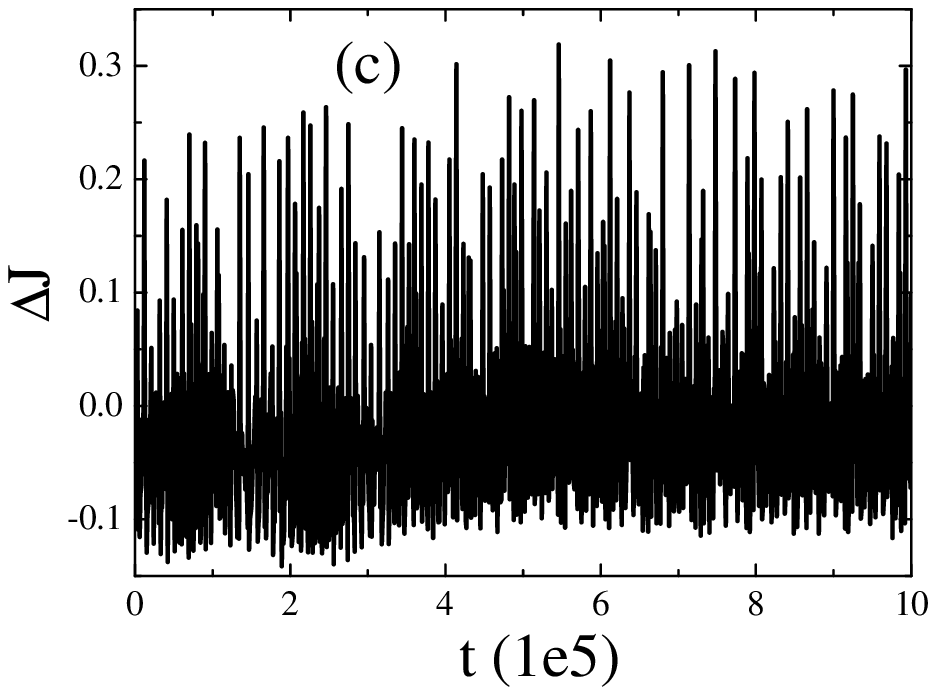}
\includegraphics[scale=0.65]{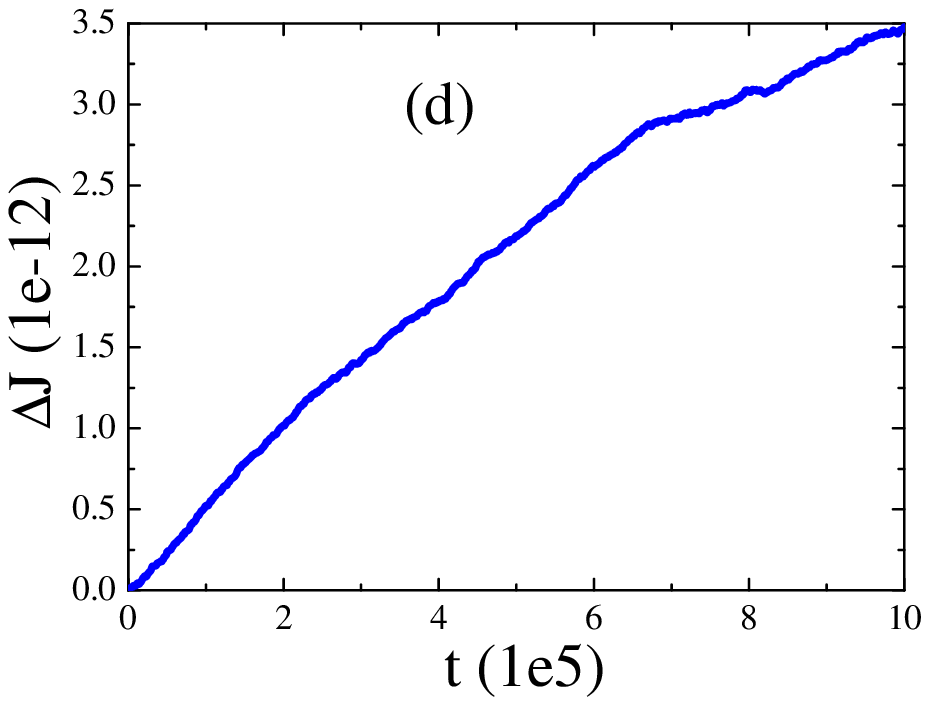}
\caption{The PN Lagrangian formulation of spinning compact
binaries: the energy errors $\Delta E$ in the incoherent equations
of motion (a) and in the coherent ones (b); the errors of the
total angular momentum $\Delta J =J_0-J$ with $J=|\mathbf{J}|$ in
the incoherent equations of motion (c) and in the coherent ones
(d). }} \label{fig5}
\end{figure*}

\begin{figure*}[tbp]
\center{
\includegraphics[scale=0.3]{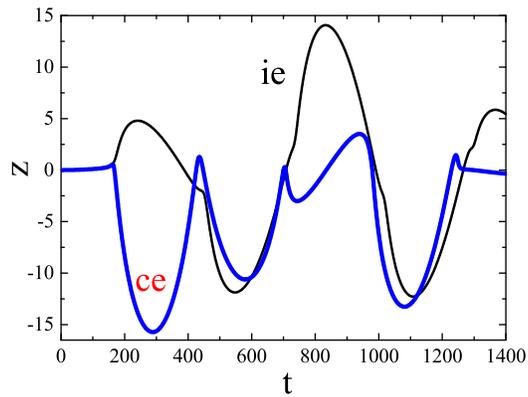}
\caption{The PN Lagrangian formulation of spinning compact
binaries: the precession values of $z$ in the incoherent equations
of motion (ie) and in the coherent ones (ce). }} \label{fig6}
\end{figure*}

\begin{figure*}[tbp]
\center{
\includegraphics[scale=0.3]{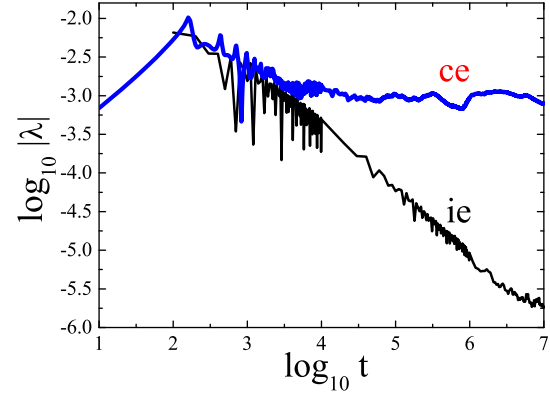}
\caption{The PN Lagrangian formulation of spinning compact
binaries: the Lyapunov exponents $\lambda$ in the incoherent
equations of motion (ie) and in the coherent ones (ce). }}
\label{fig7}
\end{figure*}

\begin{figure*}[tbp]
\center{
\includegraphics[scale=0.3]{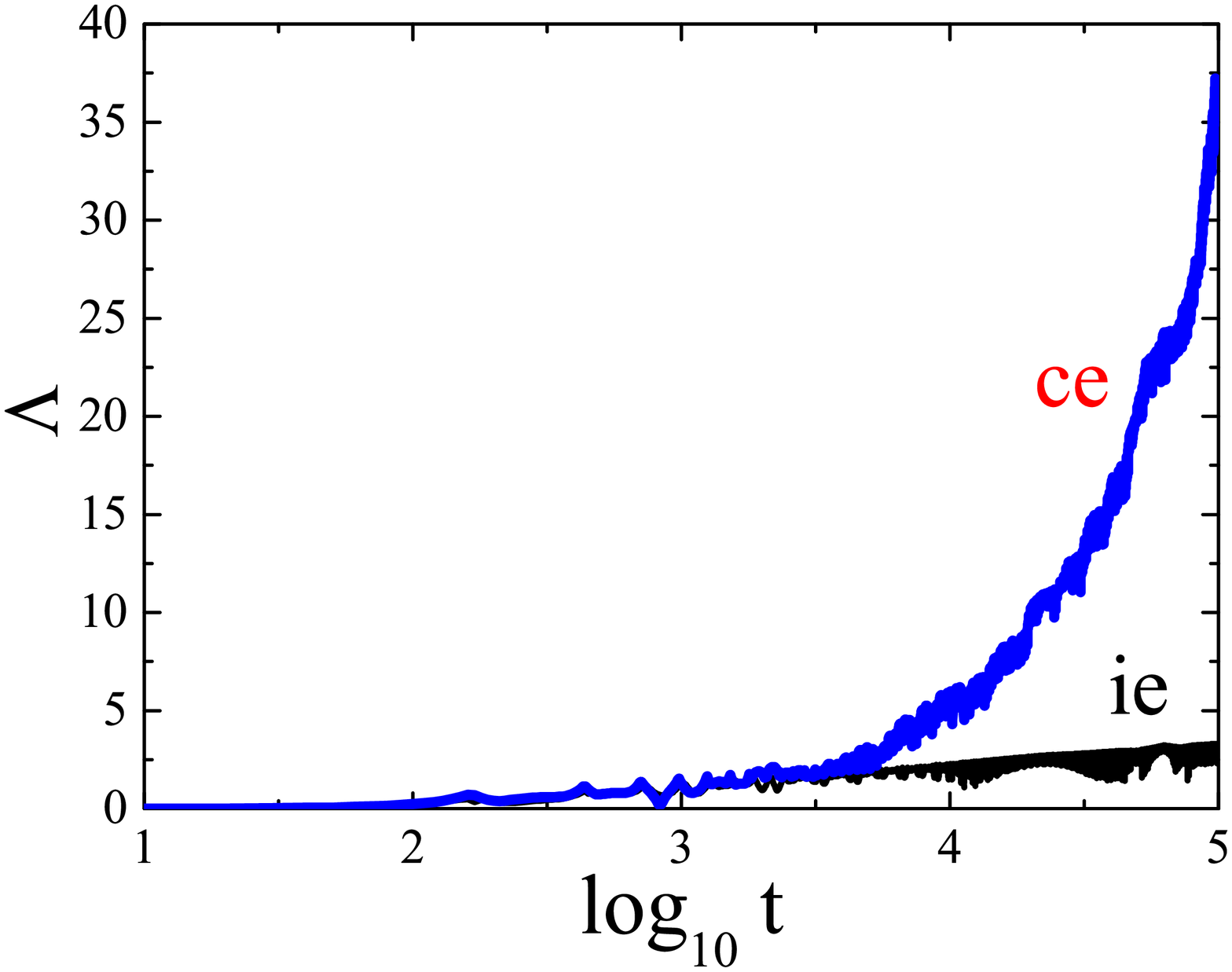}
\caption{The PN Lagrangian formulation of spinning compact
binaries: the fast Lyapunov indicators $\Lambda$ in the incoherent
equations of motion (ie) and in the coherent ones (ce). }}
\label{fig8}
\end{figure*}

\end{document}